\documentclass[conference]{IEEEtran}
\usepackage{amsmath}
\usepackage{amsfonts,amssymb}
\usepackage{array}
\usepackage{algorithm}
\usepackage{algorithmic}
\usepackage{ntheorem}
\usepackage{url}
\usepackage{subfigure}
\usepackage{url}
\usepackage[pdftex]{graphicx}

\begin{document}

\newtheorem{definition}{Definition}%[chapter]
\newtheorem{theorem}{Theorem}%[chapter]
\newcommand{\bottom}{\perp}
\newcommand{\suchthat}{{\scriptstyle\ \bullet\ }}
\newcommand{\sem}[1]{[\![#1]\!]}
\newcommand{\fdsem}[1]{[\![#1]\!]_{_{FD}}}
\newcommand{\etal}{\textit{et al.}}
\newcommand{\powerset}{\mathcal{P}}
\newcommand{\boolexp}[1]{\mathbb{B}(#1)}
\newcommand{\reduc}[1]{\!\phantom{.}_{|#1}}
\newcommand{\obs}{\otimes\!\!>}
\let\union\cup
\let\inter\cap

\title{Verification for Reliable Product Lines}
\def\email#1{Email: #1}
\IEEEoverridecommandlockouts % for \thanks-- and I had to add the mark.
\author{
\IEEEauthorblockN{Maxime Cordy\IEEEauthorrefmark{1}\thanks{\IEEEauthorrefmark{1} FNRS Research Fellow}}
\IEEEauthorblockN{Patrick Heymans}
\IEEEauthorblockN{Pierre-Yves Schobbens}
       \IEEEauthorblockA{PReCISE Research Center\\
             University of Namur, Belgium\\
            \email{mcr@info.fundp.ac.be}}
\and
\IEEEauthorblockN{Amir Molzam Sharifloo}
\IEEEauthorblockN{Carlo Ghezzi}\\
       \IEEEauthorblockA{DeepSE Research Group\\
             Politecnico di Milano, Italy\\
       \email{molzam@elet.polimi.it}}
\and
\IEEEauthorblockN{Axel Legay}\\ \\ 
       \IEEEauthorblockA{INRIA Rennes, France\\
               \email{axel.legay@inria.fr}}

}

\maketitle

\begin{abstract}
Many product lines are critical, and therefore reliability is a vital part of their requirements.
Reliability is a probabilistic property.
We therefore propose a model for feature-aware discrete-time Markov chains
as a basis for verifying  probabilistic properties of product lines, including reliability.
We compare three verification techniques:
The enumerative  technique uses PRISM, a state-of-the-art symbolic probabilistic model checker, on each product.
The parametric technique exploits our recent advances in parametric model checking.
Finally, we propose a new bounded technique that 
performs a single bounded verification for the whole product line, and thus takes advantage of the common behaviours of the product line.
Experimental results confirm the advantages of the last two techniques.

% Non-functional requirements such as reliability are of utmost importance in modern software-intensive systems. Verifying these properties on Software Product Lines (SPLs) is a challenge still to be addressed. In this paper, we propose a general framework for modelling stochasticity in SPLs. Based on this framework, we introduce a feature-aware extension of discrete-time Markov chains, aka FDTMC. We show that when randomness originates only from the environment, in which the software is embedded, an FDTMC can be decomposed into one Featured Transition Systems that models variability within the system behavior, and discrete-time Markov chains that specify the random events that impact the system. To verify properties on FDTMC, we exploit the application of recent advances in the areas of parametric model checking and SPL model checking. We present experimental results that compare these two verification approaches with an enumerative verification of each individual product, as supported by conventional model checking. The results confirm the advantages of the proposed approaches.
\end{abstract}

%\category{D.2.4}{Software Engineering}{Software/Program Verification}[Model checking]
%\category{G.3}{Mathematics of Computing}{Probability and Statistics}
%\terms{Theory, Verification}

%\keywords{~Non-Functional Requirements, Software Product Lines, Mar\-kov models, Model Checking, Probability}

%------------------------------------------------------------------------- 
\section{Introduction}
\label{sec:Introduction}

Software Product Line Engineering (SPLE) aims at developing a large number of software systems that share a common and managed set of features. In the past years, it has been an active area in both research and industry. SPLE aims at improving productivity and reducing the time, effort and cost required to develop a family of products (also called \emph{variants}). The key point to achieve this goal is to manage the variability among various products of a Software Product Line (SPL). SPLE mainly relies on model-based techniques by which variable features and behaviours are specified. The models are used to derive numerous products, each of which contains a specific set of features.

Each product of the SPL has to satisfy its functional and non-functional requirements. A common approach to assess these requirements consists in building the system and testing it. %carrying out analyses based on simulation and testing. 
However, if the system fails to meet the requirements, costly iterations are needed to improve it. This problem is even more crucial in SPLE, where a huge number of products have to be designed. Analysis techniques for single systems therefore are expensive to apply. This is the reason why researchers have recently focused on designing new quality assurance techniques dedicated to SPLs. In particular, model checking is an automated verification technique that systematically explores the whole state space of a model in search of errors. In the last years, several approaches lifted model checking to SPLs \cite{Classen2013,Cordy2013} (see more in Section~\ref{sec:related-work}). However, most of them consider qualitative properties on the sequencing of events only. They ignore important non-functional aspects of systems like reliability, availability, performance, and resource usage.

Moreover, today's software are embedded in a wide variety of systems like networks and aircrafts, that run in environments where uncontrolled events may occur randomly and affect the system. For example, a TCP transmission might be perturbed by hardware failures in a network node; a pump motor has to run faster when the flow of water goes beyond a certain threshold.
Further, some systems use randomness for their own functioning, like the USB protocol.
Markov models are widespread formalisms to model probabilistic properties of systems whose behaviour is driven by random events. Still, they cannot cope with variability. A possible approach to modeling stochastic product lines is to specify them with parametric equations; resolving these equations can be difficult, however. There is thus a need for (1) behavioural models able to represent both variability and randomness, and (2) automated techniques to efficiently verify all the modelled products against probabilistic properties.

In this paper, we propose an approach to model and verify fully observable stochastic SPLs. We enrich Markov  chains with an explicit notion of variability. This allows us to derive a new formalism that combine representations of SPL behaviour with Markov models. In particular, we introduce \emph{Featured Discrete-Time Markov Chains} (FDTMCs), an extension of discrete-time Mar\-kov chains that allows to describe a full product line in a single model, thus sharing the common parts. In FDTMCs, transitions are associated with a \emph{probability profile} that, given a product, returns the probability that the transition is executed in this product. The definition of probability profile is intended to be flexible, and allows features to arbitrarily modify the transition probability.

Next, we present and discuss three verification methods to determine which products modelled in an FDTMC (fail to) satisfy a given non-functional property. We focus more particularly on properties such as reliability that can be expressed in Probabilistic Computation Tree Logic (PCTL)~\cite{Baier2007}. A first solution to achieve this goal is to derive, for each product, the discrete-time Markov chain modelling this product from the FDTMC and to apply single-system algorithms to model check it. This \emph{enumerative} technique thus performs \emph{one exploration per product}. 
We also propose to reduce FDTMC model checking to parametric Markov-model checking, where the parameters encode the features. We can then feed the parametric model into a parametric model checker, which, in \emph{a single exploration}, yields a parametric expression that encodes the value of the desired properties. Then, \emph{for each product}, the checker has to evaluate this expression by replacing the parameters by the values corresponding to the features of the products. Our last proposition is a novel algorithm that exploits probability profiles to determine all the products satisfying the property \emph{in a single exploration}, which benefits from the compact structure of FDTMC to reduce the cost of verification. This approach applies a bounded search through the state space of an FDTMC model, and is able to calculate approximative results considering the satisfaction of a given property within a desired precision. 

To evaluate the efficiency of the three approaches, we carried out two series of experiments based on systematically extended technical examples. The results show that the enumerative approach is inefficient compared to the other two techniques, especially as the number of products grows.

%suggest that applications of the enumerative method on the whole SPLs should be avoided, and that the relative efficiency of the other two algorithms varies, depending on the number of features. %Overall, we have paved the way for the rigorous design and the efficient verification of non-functional properties in stochastic SPLs.

The remainder of the paper is structured as follows. Section \ref{PreliminaryConcepts} recapitulates background on SPL model checking. %Section \ref{MotivatingRunningExample} presents the running example used through the paper, and 
In Section \ref{sec:probability}, we recap probability theory and give its featured version. In Section~\ref{sec:dtmc}, we introduce FDTMCs, FMDP, and present our compositional modelling approach. The three verification techniques are discussed in Section~\ref{sec:verification}. Section~\ref{sec:validation} provides experimental results. We overview related work in Section \ref{sec:related-work}, and sketch future work in Section \ref{Conclusion}.

\section{Foundations}
\label{PreliminaryConcepts}

Variability in SPLs is commonly captured into features ~\cite{Kang1990}. A feature (or \emph{variation point} \cite{Pohl2005}) is a unit of difference that can differentiate two variants of an SPL. It can model optional components, functionalities of the system, but also cross-cutting behaviour. Dependencies between features may exist, and these are commonly modeled in a \emph{feature diagram}~(FD)~\cite{Kang1990}. Many different languages are used now for this purpose ~\cite{Schobbens2006}.
A product, sometimes called \emph{configuration}, \emph{variant} \cite{Pohl2005}, or \emph{model} \cite{Goguen,Kang1990},  gives a value to each feature of its FD.
A product $p$ is \emph{valid for $d$}, noted $p \vDash d$, if the values of its features satisfy all the constraints expressed in the FD. In this paper, we abstract from the syntax of FDs and define directly the semantics of an FD $d$ as its set of features (its \emph{signature} \cite{Goguen}) $\Sigma_d$ and the set of its valid products (its \emph{product line}~\cite{Schobbens2006}) $\sem{d}$. 
%This set of products will be called the described by $d$.
We can restrict a product $p$  to a sub-signature $\Sigma'$, noted $p\vert_{\Sigma'}$, which selects only the value of the features of $\Sigma'$. In other words, FD form an institution \cite{Goguen}.  %We thus stick to the semantics of Schobbens \textit{et al.}}. 

As an increasing number of SPLs are developed, quality assurance techniques for them become vital and are actively studied. FDs being unable to express behaviour, several approaches for SPL model checking have emerged during the recent years (see more in Section~\ref{sec:related-work}). Here
 we follow the principles of Featured Transition Systems (FTS)~\cite{Classen2010}. 
Assume we have an adequate model type for single products; then the semantics of the ``featured'' variant of this model is a function giving for each  product
of the product line, the corresponding model.
For instance, Transition Systems (TS) are a well-accepted model type for the qualitative behaviour of a system, and thus the semantics of FTS maps each product to a TS.
At the syntactic level, the dependance on the products is moved as much as possible inside the model, so that the common parts can be described only once.
%The main strength of FTS lies in the use of an explicit notion of features, which allows one to relate errors and undesired behaviours to the exact set of products able to produce them. 
Thus, an FTS is a state machine where transitions are annotated with \emph{feature expressions}, \textit{i.e.} formulae defined over the features. Formally, an FTS is a tuple $F = $($S$, $s_0$, $Act$, $trans$, $AP$, $L$, $d$, $\gamma$) where $S$ is a set of states, $s_0$ is the initial state, $Act$ is a set of actions, $trans \subseteq S \times Act \times S$ is a set of transitions, $AP$ is a set of atomic propositions, $L : S \rightarrow 2^{AP}$ labels a state with the propositions that it satisfies, $d$ is a feature diagram and $\gamma : trans \rightarrow (\sem{d} \rightarrow \{\top, \bottom\})$ associates a transition with a \emph{feature expression}, \textit{i.e.} a formula that encodes the set of products able to execute the transition.
The dependance on the products has thus been moved inside, to the transitions.

An FTS model-checking algorithm takes the feature information into account while looking for errors. It is thus able to keep track of the products able to execute the behaviour currently analysed, and will only examine once the common parts.
Feature expressions constitute an intuitive and flexible way to represent variability inside behavioural models. In this work, we will reuse the principles of such encoding for modelling behavioural variability in stochastic systems.

\begin{figure}
\begin{center}
\includegraphics[width=0.98\linewidth]{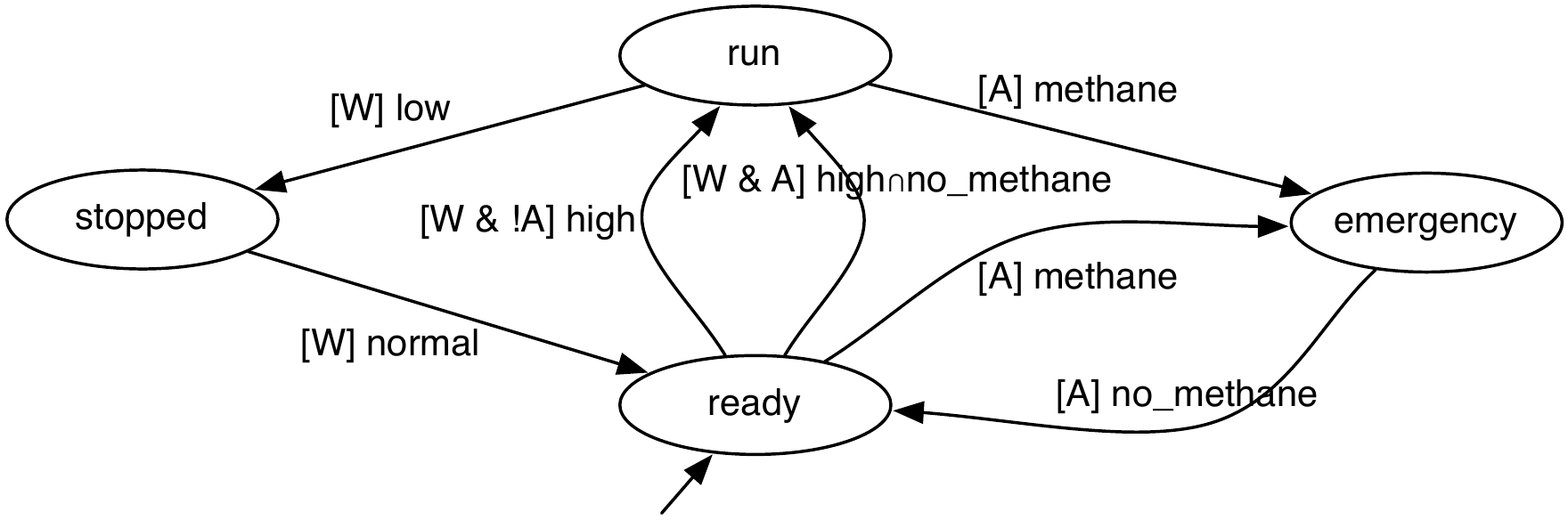}
\caption{Mine pump controller FTS}
\label{fig:fts}
\end{center}
\end{figure}

\textbf{Example}. We illustrate the concept of FTS by means of the mine pump controller~\cite{Kramer1983,Classen2012}. The objective of this software is to control a pump that drains water from a mine. When the level of water goes beyond a certain threshold, it should turn the pump on, except when there is methane in the mine, since the motor might cause an explosion. 
%In this case, the pump cannot run at all. %% false in the variant wo meth sensor
Figure~\ref{fig:fts} presents an FTS modelling the (simplified) controller SPL. Two features influence the behaviour of the controller: the presence of a \texttt{WaterSensor} ($W$), which reports the water level, and \texttt{MethaneAlarm} ($A$), which detects the presence or absence of methane. The controller starts in its state \texttt{ready}. Then, it can reach the state \texttt{run} %if action \texttt{run} is executed, typically 
when the level of water is so high that the pump has to be turned on. As specified by the associated feature expression, a product $p$ can execute this transition $t$ iff $\gamma(t)(p) = \top$, that is, iff feature $W$ is enabled in $p$. Whenever the level of water is low or there is methane, the pump must be turned off. The controller reacts accordingly and reaches state \texttt{stopped}, respectively \texttt{emergency}, provided that the controller is equipped with feature $W$, respectively $A$.

Therefore, the behaviour of the controller is determined not only by its features, but also by uncontrolled events (that is, water level and methane presence) which occur randomly. Even though FTSs are convenient for modelling behavioural variability, they cannot represent randomness. On the contrary, stochastic models cannot cope with variability. Both abilities are required to model and check non-functional quantitative properties like ``what is the probability that eventually the pump is not running even though the water level is high?'' on all the products of an SPL. The purpose of this paper is to provide a formal framework for extending stochastic models with variability information.

\section{Featured Markov Models}
\label{sec:probability}

%Nowadays, the verification of non-functional requirements like reliability and energy consumption are an important challenge in software engineering. When combined with uncertainty (that is, randomness), their assessment becomes even more cumbersome. 
Markov processes are meant to model systems where random events occur. They have been used to analyse quantitative properties in a variety of areas including economics, networks, and software. %They are, however, unable to cope with variability. 
%Here, %we revisit the basic theory of and 
%add a new dimension to Markov processes, \textit{viz.} variability. %In our approach, we assume that only the internal behaviour of the system is subject to variability, whereas randomness originates from another entity, the environment, that interacts with the system. As we will see, this assumptions allows for a separate modelling of variability and probability. 
Throughout the section, we recapitulate the fundamental definitions related to Markov processes, and enrich them with an explicit notion of variability along the principles sketched above.

Markov processes, and more generally any kind of stochastic process, are defined over a probability space and a measurable space. A probability space is a triple $(\Omega, F, P)$ where $\Omega$ is a finite set of outcomes, $F \subseteq 2^\Omega$ is a set of events, each of which is a set of outcomes,
that forms a $\sigma$-algebra, i.e., $F$ is non-empty and closed under complementation and countable unions.
 $P: F \to [0,1]$ is a probability measure function that assigns probability to events. 
The probability of a countable union of disjoint sets must be the sum of their probabilities,
and the probability of $\Omega$ must be $1$.
%Note that this notion of random events should not be confused with its common interpretation in model checking. In the latter, an event refers to an action executed by the environment that triggers a transition in the system. In our context, these actions are the outcomes of the random events. 
In our minepump example, the outcomes are the combination of water level (\texttt{low}, \texttt{normal}, and \texttt{high}) with the presence or absence of methane (\texttt{methane} and \texttt{no\_methane}, respectively); an example of event is ``there is methane and the water level is not low'', that is,  $\{(high,methane), (normal,methane)\}$. 
The state will be represented by a measurable space: it  is a couple $(S, \Sigma)$ where $S$ is a set and $\Sigma \subseteq 2^S$ is a  a $\sigma$-algebra. $S$ is called the \emph{state space}. 

Let $T$ be a totally ordered set, called \emph{time}. A stochastic process is a set $\{X_t \in S \vert t \in T\}$ of indexed, $S$-valued \emph{random variables}. Random variables are functions $X : \Omega \rightarrow S$ that maps an outcome to a state. $X_t$ represents the state of the process at time $t$.
% we stick to that informal interpretation in the rest of the paper.
 Given that we consider SPLs instead of single systems, we define that this state also depends on the features of the process. For example, if the outcome $(high,methane)$ occurs while the controller is in state ready, it should reach state \texttt{run} if it has feature $W$ or state \texttt{emergency} if it has feature $A$; otherwise it remains in state \emph{ready}. The value of $X_t$ is thus impacted by variability. Accordingly, we revise the definition of random variable, and define it as a function $X : \Omega \rightarrow (\sem{d} \rightarrow S)$ that, given an outcome $\omega$ and a variant $p$, returns the state reached by $p$ following the occurrence of $\omega$. This revised definition leads us to the following extended definition of stochastic process.
\begin{definition}
Let $(\Omega, F, P)$ be a probability space, $(S,\Sigma)$ be a measurable space, $T$ a totally ordered set, and $d$ a feature diagram. A featured stochastic process is a set $\{X_t \in S^{\sem{d}} \vert t \in T\}$ where for any $t$, $X_t$ is a random variable.
\end{definition}
According to this definition, the state reached by the process at time $t$ is determined by (1) the outcome and (2) the product, i.e. the combination of features of the process. 

% TODO: Move to outlook?/conclusion
%Our definitions capture the variability of the system only. When the environment is not static (\textit{e.g.}, in the case of adaptive systems~\cite{Cordy2013b}), some approaches postulate that features may capture its dynamic characteristics and associate them to the set of actions the environment may perform on the system. Our generalisation of Markov processes implicitly consider these cases. Indeed, modifications in the environment features can be regarded as outcomes defined in the probability space, and thus may affect the value of the random variables that defines the states reached by the system.

Markov processes are a particular type of stochastic processes that satisfy the so-called Markov property. Intuitively, this property, often referred to as the \emph{memoryless} property, implies that the probability of reaching a state at a future point of time can be determined knowing only its current state. Since we revised the definition of random variables and stochastic process, we extend the definition of the Markov property as well, and obviously we require that the property holds for each product.
\begin{definition}
A featured Markov process is a varia\-bility-aware stochastic process $\{X_t \in S^{\sem{d}} \vert t \in T\}$ such that $\forall p \in \sem{d} \suchthat \forall t_1 < t_2 \in T \suchthat s \in S$ we have
\begin{multline*}
Pr(X_{t_2}(p) = s ~\vert~ X_u(p) = s_u, \forall u \leq t_1) \\ = Pr(X_{t_2}(p) = s \vert X_{t_1}(p) = s_{t_1})
\end{multline*}
\end{definition} 
Assume furthermore that time $T$ is equipped with an addition.
\begin{definition}
A featured Markov process is time-homogeneous if %and only if
$$Pr(X_{t_1+t_2}(p) = s \vert X_{t_1}(p) = s') = Pr(X_{t_2}(p) = s \vert X_{0}(p) = s')$$
\end{definition} 
When the time is discrete, it is enough to take $t_2=1$.

There exist many types of Markov models: First, we can model time as continuous ($T = {\mathbb R}_{\ge 0}$) or discrete ($T = {\mathbb N}$). 
Second, we can assume that the process is purely stochastic (Markov chain)
or partially under control of an opponent (Markov decision process).
Third, rewards can be added to the model (Markov reward model), etc. 
Specialized algorithms have been developed for each type of model \cite{BHHK03,CKKP05}.
In the rest of this paper, we will focus on the two most widely used models, namely time-homogeneous 
discrete-time Markov chains (DTMC) and time-homogeneous discrete-time Markov decision process (MDP).
For each, we will derive its featured extension (FDTMC and FMDP).
We leave for future work the algorithmic treatment of other featured stochastic processes, but their
semantics is already defined in this section.

\section{Featured Discrete-Time Markov Processes}
\label{sec:dtmc}

The above definition of featured Markov processes is the semantic foundation for the definition of formalisms to model probabilistic requirements in SPLs. Here, we specialize it to Featured Discrete-Time Markov Chains (FDTMCs), a formalism that extends Discrete-Time Mar\-kov Chains (DTMCs) to allow modelling both variability and stochasticity. In this section, we first briefly illustrate how classical DTMCs model the behaviour of a stochastic environment. Then we present the syntax and the semantics of FDTMC. 
%Based on these definitions, we propose a compositional modelling approach that relies on a division of the stochasticity in the environment and the variability in the SPL. We make use of DTMC and FTS to model the environment and the SPL behaviour, respectively. Finally, we define the semantics of the composition of these two models in terms of FDTMC.

\subsection{Modelling random events with DTMCs}
DTMCs are a particular type of Markov processes where (1) the state space of the system is finite, and (2) time elapses at discrete steps. DTMCs can be regarded as transition systems where transitions are annotated with a probability value that describes the likelihood of their occurrence. These probabilities satisfy the usual probability axioms. In particular, for a given state, the probabilities of its outgoing transitions must sum to 1. 
%DTMCs are useful to model random events that occur, typically actions performed by an uncontrolled environment. 

\begin{figure}[htb]
	\begin{center}
%	\subfigure[Methane]{
		\includegraphics[width=0.6\linewidth]{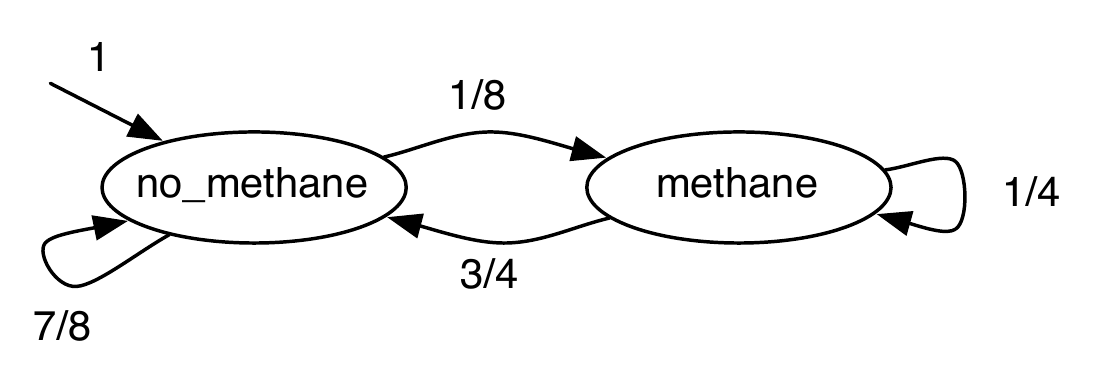}
%	}
%	\subfigure[Water level]{
%		\includegraphics[width=0.75\linewidth]{figure/minepump-water.pdf}
%	}
	\caption{The DTMC modelling methane evolution}
	\label{fig:minepump}
	\end{center}
\end{figure}

A example DTMCs is shown in Figure~\ref{fig:minepump}. It models the evolution of methane %and water
evolution in the mine in which the aforementioned pump is installed. Initially, %the water level is high and
 there is no methane. At the next discrete point of time, either there is methane (probability 0.125) or not (probability 0.875).
Then the methane, if present, will disappear spontaneously in 75\% of the cases.
Note that the probability of staying in the same state (self-loop) can be deduced from the other outgoing transitions, since they have to sum up to 1; we will thus omit them in the rest of the paper.
% and the water level either remains high (with a probability of 0.25) or becomes normal (with a probability of 0.75).
%The environment of the pump system is given by the independent product of these two DTMCs.
DTMC are adequate to model systems that evolve spontaneously, without influence from the outside world.

\subsection{FDTMC: A featured stochastic formalism}
Although DTMCs are convenient to represent stochasticity, they cannot cope with variability. To represent stochastic behaviour in SPLs, we enrich them in the same way as we enhanced transition systems with feature expressions to obtain the FTS formalism~\cite{Classen2011,Classen2013}. The original intent of feature expressions is to specify that features can add or remove transitions. In DTMCs, however, the addition or removal of a transition leaving a given state may modify the probability of its other outgoing transitions. More generally, features can modify the probability distribution of any transition. The only restriction is the satisfaction of the probability axioms in all the products. Therefore, the probability of a given transition is not a fixed value anymore. To represent this new kind of variability, we propose to annotate transitions with a \emph{probability profile} $\sem{d} \rightarrow [0,1]$, by a function $\Pi : S \times S \rightarrow (\sem{d} \rightarrow [0,1])$ that associates to each transition a a product with a probability value. Intuitively, $\Pi(s,s')(p)$ is the probability of occurrence of the transition $(s,s')$ in the product $p$. In the figures, a profile is drawn as several arrows, one per  probability value, and each arrow has a guard that describes the set of products that yield this probability value. %In reality, however, it is unlikely that all the transitions will receive a different probability value for every product. In a given model $M$, we call the probability profiles of $M$ the set of probability profiles that determine the value of the parameters. 

\begin{definition}
A Featured Discrete-Time Markov Chain (FDTMC) is a tuple $(S, s_0, d, \Pi, A, L)$ where:
\begin{itemize}
\item  $S$ is a finite, non-empty set of states;
\item $s_0 \in S$  is the initial state; % \to [0,1]$ is the initial probability distribution on states, where $\sum_{s \in S} \nu(s) = 1$;
\item $d$ is a feature diagram;
\item $\Pi : S \times S \rightarrow (\sem{d} \rightarrow [0,1])$ is the transition probability function, which assigns a probability profile to each transition. Equivalently, for each starting state and each product, it gives a probability distribution on the target states.
Any probability profile must satisfy the probability axiom for all the products:
$$ \forall p \in \sem{d}, \forall s \in S, \sum_{s' \in S} \Pi(s,s')(p) = 1$$
\item $A$ is a set of atomic predicates;
\item $L: S \to 2^{A}$ labels each state by the set of predicates that holds there.
\end{itemize}
\end{definition}
An FDTMC is a concise representation for a family of DTMCs, that is, one per valid product. The DTMC modelling a particular variant $p$ is obtained by \emph{projecting} the probability profile of each transition onto $p$. The transition probability function of the resulting DTMC is defined as $P : S \times S \rightarrow [0,1] : P(s,s') = \Pi(s,s')(p)$. It generalises DTMC since%Of course
, when $\sem d$ is a singleton, a FDTMC is simply a DTMC.
%If that is the case, the profile is said to be \emph{product-consistent}. 

All usual operations on probabilities can be extended by considering them as a function of the product line.
For instance, the product of two (independent) probability profiles $\Pi$ and $\Pi'$ is defined as $(\Pi \otimes \Pi')(p) = \Pi(p) . \Pi'(p)$.
% The sum $\oplus$ of two probability profiles is defined similarly. We denote by \textbf{0} (resp \textbf{1}) the profile that associate the value 0 (resp. 1) to every product. Then the complement of $\Pi$ is defined as $(-\Pi) : (-\Pi)(p) = 1 - \Pi(p)$, also noted $\textbf{1} - \Pi$. From the notion of probability profile results our featured extension of DTMC.

A \emph{path} of an FDTMC is a % (possibly infinite) %% def does not work for inf case.
sequence of its states. Given that transition probability depends on features, the execution probability of a path does as well. Let $s \in S$ be a state of an FDTMC, and $Paths(s)$ the set of paths starting from $s$. Due to the Markov property, the probability that a finite path $\rho =  \rho[0], \rho[1], \dots, \rho[n] \in Paths(s)$ is executed is given by the product of the probability profiles of its transitions:
$$\Pi(\rho) = \Pi(\rho[0],\rho[1]) \otimes \dots \otimes \Pi(\rho[n-1],\rho[n]).$$ 
This constitutes the base of the unique family of probability measure on paths, by the usual cylinder set construction \cite{KSK76}.
%The execution probability of a set of paths starting from a state $s$ is obtained by summing the probability of the paths contained in this set: $\Pi(\{\rho_1, \dots, \rho_m\},p) = \sum_{j = 1}^{m} \Pi(\rho_j,p)$. 
%According to this definition, the probability of any subset of $Paths(s)$ is a lower bound of $\Pi(Paths(s))$.

\begin{figure}[htb]
	\begin{center}
	\includegraphics[width=0.6\linewidth]{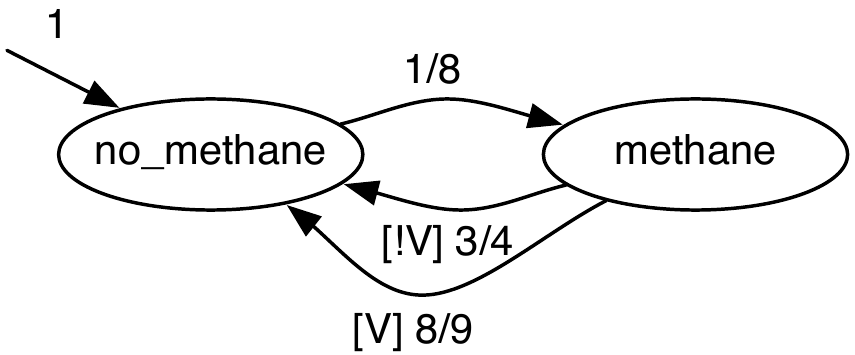}
	\caption{Methane evolution in a mine with optional ventilation}
 	\label{fig:minepump-fdtmc}
 	\end{center}
\end{figure}

For instance, the FDTMC of Figure~\ref{fig:minepump-fdtmc}  (where self-loops are omitted) represents a mine that could be equipped by a natural ventilation system,
by creating well-placed air entrances. This feature is static: it is selected at the construction of the mine.
It cannot prevent the apparition of methane, but it will help dissipating it.

% Figure~\ref{fig:minepump-fdtmc} shows an excerpt of the FDTMC modelling the minepump SPL together with its stochastic environment. It depicts the transitions leaving the FDTMC state where the controller is \emph{ready}, the water level is normal and there is no methane in the mine. In this graphical representation, we annotate every transition with expression of the form $[f]pr$ to specify that a given transition occurs with a probability $pr$ if the feature expression $f$ is satisfied; otherwise the transition probability is zero. 

FDTMC is a fundamental formalism, which is not meant to be used by engineers. Because of the flexibility of probability profiles, FDTMC may be hard to write down. Many states may be needed to entirely represent the system and its environment. Moreover, many transitions have a variable probability value that depends on the features of the system. When the number of features and transitions grows, it becomes increasingly harder to have a comprehensive view on the stochastic behaviour of every variant. Therefore we introduce ways to decompose the description. A problem that we meet is that (F)DTMC are meant to described closed systems.
Therefore we can decompose a system into  (F)DTMCs only if the subsystems are completely independent and do not communicate.
Therefore, we introduce Markov Decision Processes (MDP) as an intermediary to decompose our systems.
%Our current tools, however, are not compositional: a first tool will compose the FMDPs to create a single FDTMC,
%then another will verify the properties

\subsection{FMDP: a Composable Stochastic Model}

FMDP is a model adequate to represent composable, communicating, stochastic, featured systems.
They generalise MDP, FTS and FDTMC.
In this paper, we use FMDPs only to create the final FDTMC that will be model-checked.
However, they are of independent interest, and there are tools that deal with MDPs directly \cite{Kwiatkowska2009,Baier05}.

\begin{definition}
A Featured Discrete-Time Markov Decision Process (FMDP) is a tuple $(S, s_0, Act, d, \Pi, A, L)$ where:
\begin{itemize}
\item  $S$ is a finite, non-empty set of states;
\item $s_0  \in S $ is the initial state;
\item $Act$ is a finite set of actions;
\item $d$ is a feature diagram;
\item $P : S \times Act \times S \rightarrow (\sem{d} \rightarrow [0,1])$ is the transition probability function, which assigns a probability profile to each transition. Equivalently, for each starting state, each action and each product, it gives either a probability distribution on the target states, or always $0$ to indicate that the action is not enabled.
It must thus satisfy the consistency axiom:
\begin{align}
 \forall p \in \sem{d}, \forall s \in S, \forall a \in Act, \sum_{s' \in S} P(s,a,s')(p) = 1 \textrm{ or  }0 \label{FMDP}
\end{align}
\item $A$ is a set of atomic predicates;
\item $L: S \to 2^{A}$ labels each state by the set of predicates that hold there.
\end{itemize}
\end{definition}

When actions are always enabled, the FMDP is complete:
\begin{definition}
A FMDP is \emph{complete } if
$$ \forall p \in \sem{d}, \forall s \in S, \forall a \in Act, \sum_{s' \in S} P(s,a,s')(p) = 1$$
% for any state $s \in S$, action $a \in Act$, product $p \in \sem{d}$, there is exactly one successor state $s'$, i.e. such that $\gamma(s,a,s')(p)$ is true.  
\end{definition}

When $P$ returns either 0 or 1, a FMDP reduces to a deterministic FTS. 
When there is a unique action (that we call $tick$), a complete FMDP reduces to a FDTMC.
When $d$ has a unique product, a FMDP reduces to a MDP. 

For instance, the FMDP of Fig.~\ref{fig:minepump-fdtmc} (where self-loops are omitted) represents the evolution of the water level in the mine.
The water can raise spontaneously (due to flooding) but it can also decrease spontaneously (due to evaporation and infiltration).
Running the pump wil favour  decrease of the water level. Note that here the guards contains dynamic predicates (written in lowercase)
and not static features.

\begin{figure}[htb]
	\begin{center}
	\includegraphics[width=0.98\linewidth]{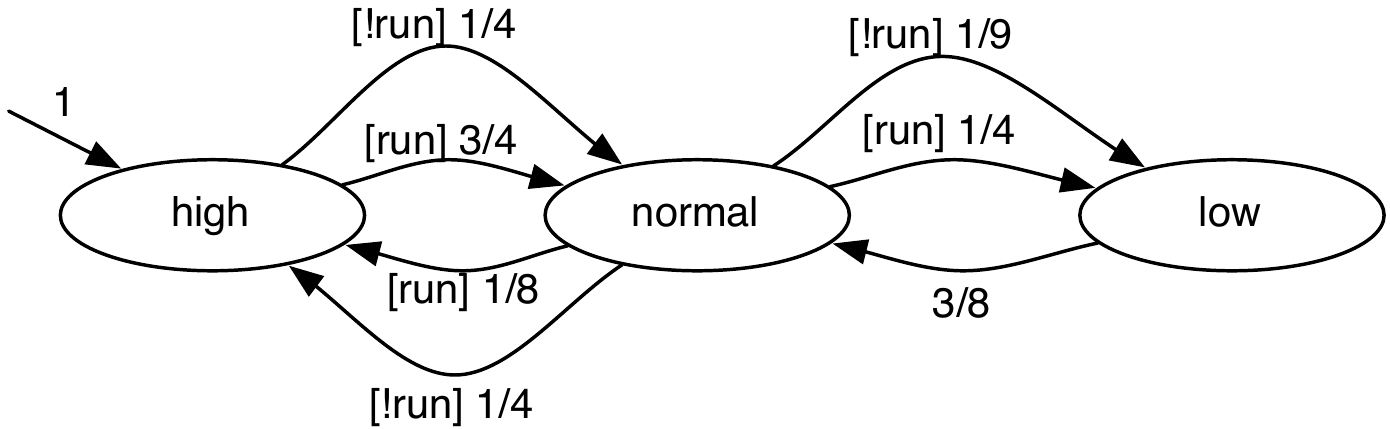}
	\caption{Water evolution in a mine with a pump}
 	\label{fig:water-fmdp}
 	\end{center}
\end{figure}

While most authors (e.g. \cite{Baier05}) give to MDP a semantics under a probabilistic infinite-memory scheduler that chooses the actions,
we will not do so here, since our goal is to eliminate this non-determinism by composing the processes.

\subsection{Composition}
%We postulate that most of software-intensive systems are intended to exhibit a deterministic inner behaviour; the sto\-chasticity thus originates only from an uncontrolled environment that behaves in a random manner. Moreover, we consider that features impact on the behaviour of the system only. Under these conditions, it is possible to decompose an FDTMC into two different automata. On the one hand, standard DTMCs describe the probability of outcomes that occur in the environment. A state of the DTMC therefore represents the outcomes of the environment, which are randomly triggered. On the other hand, Featured Transition Systems model the reactions of the system to these outcomes with respect to its features.  In this FTS, actions are replaced by events (that is, sets of outcomes), such that a transition in the FTS is executed by a product $p$ iff (1) the current outcomes of the environment are included in the event labelling the transition, and (2) the features of $p$ satisfy the feature expression labelling the transition. If no transition meets these requirements, the FTS remains in its current state.

To make FDTMC descriptions more manageable, we use classical composition operators:
\begin{itemize}
\item  the \emph{synchronized product}, borrowed from PRISM reactive modules:
actions can be synchronized in the style of CSP, and each process can read the predicates of other processes in its guards, but can only change its own state. We model these guards in the actions. 
The probabilistic choices of each component are independent.
%\item the \emph{state-synchronized product}
\item  the \emph{observer product} is asymmetric:  the second process can immediately observe the predicates of the first one, but not conversely. If we add the converse, we would create a causality loop.
The controllers are usually modelled as observers: it is assumed that their reaction is much faster than the environmental evolution, and can be considered instantaneous.
%the random choices of
% since a FDTMC models a system without inputs.
\end{itemize}

% \begin{figure}
% 	\begin{center}
% 	\includegraphics[width=0.98\linewidth]{figure/minepump-parallel.pdf}
% 	\caption{The minepump controller} % modelled according to the random events that occur in the mine.}
% 	\label{fig:minepump-parallel}
% 	\end{center}
% \end{figure}

For instance, the FDTMC of the mine pump %of Figure~\ref{fig:minepump-fdtmc} 
can be separated into an FDTMC (i.e. without inputs) for methane (Figure~\ref{fig:minepump-fdtmc}), an MDP (i.e. without features) for water (Figure~\ref{fig:water-fmdp}), and an FTS (i.e. without probabilities) for the controller (Figure~\ref{fig:fts}).
%shown in Figures~\ref{fig:minepump} and~\ref{fig:minepump-parallel}, respectively.
% The two DTMCs model the level of water and the presence of methane, whereas the FTS model the controller of the system that describes its reactions with respect to the outcome of the two stochastic processes.
Let us assume that, in a system without ventilation ($\neg V$), % but with methane and water sensors ($A \land V$) 
there is \texttt{no\_methane} and the water level is \texttt{normal}, whereas the controller is \texttt{ready}. At the next discrete point of time, the probability to reach state with  \texttt{methane} and \texttt{high} water,  %respectively,
 is given by $\frac{1}{8}.\frac{1}{4} = \frac{1}{32}$. In the FTS, there are two outgoing transitions of \texttt{ready} that are labelled by an event including this outcome: one leads to \texttt{run} and is labelled by event \texttt{high} %$High$
 and feature expression $W \land \lnot A$; the other leads to \texttt{emergency} and is labelled by event \texttt{methane} and feature expression $A$. Hence, the system will reach state \texttt{run} if it is equipped with feature $W$ but not $A$, will reach state \texttt{emergency} if feature $A$ is enabled, and will stay in state \texttt{ready} otherwise.

When composing featured systems, we must also compose their FD. We use the operator $d_1 \land d_2$ defined by 
$\sem{d_1 \land d_2} = \{ p \in Mod(\Sigma_1 \union \Sigma_2) \textrm{ such that }  p|_{\Sigma_1} \in \sem{d_1},   p|_{\Sigma_2} \in \sem{d_2} \}$.
%Its signature is the union of the signatures of $d_1, d_2$.
\goodbreak
\begin{definition}
  The \emph{synchronized product} %on state and common actions in $Act_1 \inter Act_2$ 
of two FMDPs $M_1 = (S_1, i_1, Act_1 \times 2^{A_2},$  $ d_1, P_1, A_1, L_1)$ and $M_2 = (S_2, i_2, Act_2 \times 2^{A_1},$ $ d_2, P_2, A_2, L_2)$ (with $A_1, A_2$ disjoint) is the  FMDP $M = M_1 \| M_2$ given by:
  \begin{itemize}
  \item $S = S_1 \times S_2$
  \item $s_0 = (i_1,i_2)$
  \item $Act = Act_1 \union Act_2$
  \item $d = d_1 \land d_2$ % = \{ p \in Mod(\Sigma_1 \union \Sigma_2) |   p|_{\Sigma_1} \in \sem{d_1},   p|_{\Sigma_2} \in \sem{d_2} \}$
  \item $A = A_1 \union A_2$
  \item $L((s_1,s_2)) = L_1(s_1) \union L_2(s_2)$
  \end{itemize}
 If $a \in Act_1 \inter Act_2$:
                 $P((s_1,s_2),a,(s_1',s_2'))(p)$ 
                        $$= P_1(s_1,(a,,L_2(s_2)), s_1')(p|_{\Sigma_1}) .  P_2(s_2,(a,L_1(s_1)),s_2')(p|_{\Sigma_2})$$ 
             If $a \in Act_1 \setminus Act_2$:
                $P((s_1,s_2),a,(s_1',s_2'))(p)$ 
                        $$= P_1(s_1,(a,L_2(s_2)),s_1')(p|_{\Sigma_1}) . (s_2= s_2')$$
           If $a \in Act_2 \setminus Act_1$: 
                $P((s_1,s_2),a,(s_1',s_2'))(p)$  
                     $$ =  (s_1= s_1') . P_2(s_2,(a,,L_1(s_1)),s_2')(p|_{\Sigma_2})$$ 
\end{definition}
Above, $(s=s')$ is a function that returns 1 when $s,s'$ are equal, and 0 otherwise.
\begin{theorem} 
If $M_1, M_2$ are complete FMDPs, then their synchronized product $M_1 \| M_2$ is also a complete FMDP.
\end{theorem}

This definition can also be used between two FDTMC, by considering them as single-action complete FMDP.
Then, they synchronize on their common unique action,
and their synchronized product represents their execution, synchronized on time steps, but probabilistically independent.
The theorem above shows that the result is again  a single-action complete FMDP, i.e. a FDTMC.

The product of FMDPs without shared actions represents their interleaved execution.

% Often, one process needs to base its choices on the states of the others. To this end, we provide the state-synchronized operator:
% \begin{definition}
%   The \emph{product} synchronized states of two FMDPs $M_1 = (S_1, i_1, Act_1 \times 2^{A_2}, d_1, P_1, A_1, L_1)$ and $M_2 = (S_2, i_2, Act_2 \times 2^{A_1}, d_2, P_2, A_2, L_2)$ (with $A_1, A_2$ disjoint) is  $M = M_1 \|_S M_2$ given by:
%   \begin{itemize}
%   \item $S = S_1 \times S_2$
%   \item $s_0 = (i_1,i_2)$
%   \item $d = d_1 \land d_2$ %\{ p \in Mod(\Sigma_1 \union \Sigma_2) |   p|_{\Sigma_1} \in \sem{d_1},   p|_{\Sigma_2} \in \sem{d_2} \}$
%   \item $P((s_1,s_2), (a_1, a_2), (s_1',s_2')) = P(s_1, (a_1, L_2(s_2)), s_1') \otimes P(s_2, (a_2, L_1(s_1)), s_2')$
% \let\union\cup
%   \item $A = A_1 \union A_2$
%   \item $L((s_1,s_2)) = L_1(s_1) \union L_2(s_2)$
%   \end{itemize}
% \end{definition}

%Another product is often needed.
%To simplify the upcoming definitions, we assume that in the FTS, there exists at most one transition from a state $s_f$ to another state $s_f'$; we denote by $e(s_f,s_f')$ the event labelling this transition. This assumption does not decrease the expressiveness of our compositional formalism: if there exists such two transitions with events $e_1(s_f,s_f')$ and $e_2(s_f$,$s_f')$, we can merge them into one transition labelled with event $e_1(s_f,s_f') \cup e_2(s_f,s_f')$. 
In the \emph{observer product}, the actions that drive the observer will again be %boolean combinations 
the sets of the atomic predicates of the observee, but they are now read immediately (hence the prime in the definition of transitions).
% In the figures, this is indicated by a prime after the name of the predicate.
%in the next state.
%The observee will evolve on its own: therefore we require the observer to be ready to accept any input from the observee.

%The semantics of this composition is defined as the FDTMC resulting from the \emph{synchronous product} of a DTMC with an FTS.
\begin{definition}
The \emph{observer product} of a FMDP $M$ = $(S_1, i_1, Act,$ $ d_1, P_1, A_1, L_1)$ %defined over a probability space $(\Omega, E, P)$ %%never used
with a FMDP  $M_2$ = $(S_2, i_2, 2^{A_1}, d_2, P_2, A_2, L_2)$  (with $A_1, A_2$ disjoint) is  $M = M_1 \obs M_2$ given by:
%FTS $F = (S_2, s_0, 2^{A_1}, \rightarrow, A_2, L_2, d_2, \gamma)$ is the FDTMC $M \time
\begin{itemize}
\item $S = S_1 \times S_2$;
\item $i =(i_1,i_2)$ %= \vu_1(s_1)$;
%$\vu((s_1,s_2)) = 0$ when $s_2 \neq s_0$.
\item $ Act = Act_1$ % = 2^{A_1}
\item $d = d_1 \land d_2$ %$\sem{d} = \{ p \in Mod(\Sigma_1 \union \Sigma_2) |   p|_{\Sigma_1} \in \sem{d_1},   p|_{\Sigma_2} \in \sem{d_2} \}$
\item 
 %(, \nu_1, d, \Pi)$ where, for $(s_1, s_f), (s_1', s_f') \in S_1 \times S_f$:
%\begin{align*}
%\nu(s_1,s_f) &= \left \lbrace \begin{array}{lr}
%\nu_1(s_1), & s_f = s_0;\\
%0, & \texttt{otherwise.}
%\end{array} \right . \\
 %&= \left \lbrace \begin{array}{lr}
$P((s_1,s_2),a,(s_1',s_2'))(p)$
   $$= P_1(s_1,a,s_1')(p|_{\Sigma_1}) . P_2(s_2,L_1(s_1'), s_2') (p|_{\Sigma_2})$$
 % &  \gamma(s_2,L_1(s_1),s_2')(p|_\Sigma_2)\\
%0, & \texttt{otherwise.}
%\end{array} \right .
%\end{align*}
  \item $A = A_1 \union A_2$
  \item $L((s_1,s_2)) = L_1(s_1) \union L_2(s_2)$
  \end{itemize}
\end{definition}

Note that the definition above does not always yield a proper FMDP: the probabilities could sum to a number stricly between 0 and 1.
%As mentioned before, the probability profile of the FDTMC must be product-consistent. A sufficient condition is that the FMDP is complete:
% drawback : impossible to model when the possible outcomes depend on the system states
\begin{theorem} % was false
Let $M_1 , M_2$ be FMDPs with $Act_2 = 2^{A_1}$ as above. %defined over a probability space $(\Omega, E, P)$ 
Then,  $M_1 \obs M_2$ is consistent for axiom (\ref{FMDP}) if $M_2$ is complete.
%for any product $p \in \sem{d}$ and FTS state $s_f \in S_f$, we have:
%\begin{align*}
% \forall (s_f', s_f'') \in S_f \times S_f \setminus \{s_f'\} & \suchthat e(s_f,s_f') \cap e(s_f,s_f'') = \emptyset.
%\bigcup_{s_f' \in S_f} e(s_f,s_f') &= S_d
%\end{align*}
\end{theorem}

Further, we have:
\begin{theorem}
Let $M_1 , M_2$ be FMDPs with $Act_2 = 2^{A_1}$ as above. %defined over a probability space $(\Omega, E, P)$ 
Then,  $M_1 \obs M_2$ is a complete FMDP  if $M_1, M_2$ are complete.
%for any product $p \in \sem{d}$ and FTS state $s_f \in S_f$, we have:
%\begin{align*}
% \forall (s_f', s_f'') \in S_f \times S_f \setminus \{s_f'\} & \suchthat e(s_f,s_f') \cap e(s_f,s_f'') = \emptyset.
%\bigcup_{s_f' \in S_f} e(s_f,s_f') &= S_d
%\end{align*}
\end{theorem}
In particular, a FDTMC observed by a complete FMDP is again a FDTMC.

These operators are only a basis for a usable language. We plan to unify fPromela \cite{Classen2012} and Probmela~\cite{Baier05}
to obtain a modelling language easy to use (at least for Promela modellers). %, and containing these operators.

%Since this modelling method allows a clear separation between variability and stochasticity, it also permits to reuse existing high-level languages defined on top of FTS (\textit{e.g.}, fPromela~\cite{Classen2012})
%% does not work because fPromela has no notion of actions
% and DTMCs (\textit{e.g.}, Probmela~\cite{Kwiatkowska2009}),
%% Probmela is for MDP
% respectively. By combining these languages with our synchronous product operator, we bridge the gap between engineers and the FDTMC formalism.
The FDTMC modelling the minepump SPL is obtained by  computing the synchronized composition of the the FDTMC for methane and the FMDP for water, while the deterministic completed FTS representing the controller is composed as an observer.
%and (2) computing the observer product of the resulting DTMC with. 
The theorems above show that the result is indeed a FDTMC. It has three boolean features (\verb+W, A, V+), 8 products, and 24 states. For comparison, the Linux product line has about 10 000 features.

\section{Verification of probabilistic properties}
\label{sec:verification}
By combining Markov models with variability information, we pave the way for automating the verification of probabilistic properties in SPLs. %Still, we have to provide algorithms for checking such models. 
We also need to provide a language to express those properties; 
%focus more particularly on properties like reliability that can be expressed in
here we use PCTL~\cite{Baier2007}, that allows to express reliability properties.
We want to assess the satisfaction of such properties for all the valid products of a given SPL. An example of property to verify is ``Which products guarantee that the probability that the pump eventually runs in presence of methane, is less than 0.1?''. %Another is ``how many instances of feature \emph{working unit} are needed to ensure that the probability for a service-based architecture to be operational is higher than 0.9?``. 
PCTL formulae are defined according to the following syntax:
\begin{align}
\notag 
 \begin{split}
	& 
	\Phi::=  true \ \vert \ a \ \vert \ \Phi \ \wedge \ \Phi \ \vert \ \neg \ \Phi \ \vert \ \mathbb{P}_{J} \  (\Psi)
	\\ &
	\Psi ::= X \Phi \ | \ \Phi_{1} U^{\leq t} \Phi_{2} | \ \Phi_{1} U \Phi_{2}
\end{split}
\end{align}
where $p \in [0,1]$, $J$ is an interval $\subseteq [0,1]$, $t \in \mathbb{N} \cup \{ \infty \}$, and $a\in A$ is an atomic proposition. Formulae generated from $\Phi$ are referred to as \emph{state formulae} and since can be evaluated to either true or false in every state of a product. $\mathbb{P}$ is named the \emph{probability operator} and $\mathbb{P}_{J} \  (\Psi)$ specifies that the probability that $\Psi$ is satisfied from a given state must be within interval $J$.  Formulae generated from $\Psi$ are named \emph{path formulae} and their truth is to be evaluated for each execution path. The temporal operator $X$ is called \emph{Next}, and $X \Phi$ means that $\Phi$ must hold in the next state. $U$ is called \emph{Until}. Intuitively, a path satisfies $\Phi_1 U \Phi_2$ iff $\Phi_2$ holds in some state in the path and $\Phi_1$ holds in every preceding state. $U^{\leq t}$ is a bounded variant of the until. A path satisfies $\Phi_1 U^{\leq t} \Phi_2$ iff it satisfies $\Phi_1 U \Phi_2$ in at most $t$ steps. We propose three methods to verify FDTMC against PCTL formulae.

\subsection{Enumerative Model Checking}

Our first method, called \textbf{enumerative}, checks an FDTMC using standard DTMC verification algorithms. To that aim, it computes the projection of the probability profiles of the FDTMC to every product, and then model checks the resulting DTMCs individually. The computation of the projection requires apply the probability profile of every transition of the FDTMC. When the FDTMC is modelled as a product of components, one may instead compute the projection of each component, and then build back the same product of the resulting MDPs (without features). Further, if no features appear in a component, like the DTMC of the example,
the projection does nothing.
% In this case, only the transitions of the FTS are analysed, as opposed to the complete FDTMC. 
The correctness of this method is guaranteed by the fact that the projection operator is distributive over the synchronized and observer product.
\begin{theorem}
Let $M, N$ be FMDPs. Then $\forall p \in \sem{d},  M_{|p} \| N_{|p} = (M \| N)_{|p}$ and $M_{|p} \obs N_{|p} = (M \obs N)_{|p}$.
\end{theorem}
An undeniable advantage of the enumerative approach is the possibility to reuse the most efficient state-of-the-art tools for single-systems, with their numerous optimisations. In particular, matrix analytic methods~\cite{Latouche1999,Baier2007} cannot be applied on FDTMCs given that in these models, probability distributions cannot be represented as a two-dimensional stochastic matrix; it would be a matrix of profiles. Another advantage is that this method is appropriate for \emph{product sampling}-based verification, where only a (small) subset of the products are verified. The enumerative approach, however, does not benefit from the commonalities between the products and thus performs more redundant verifications as the number of products increases. %In consequence, we propose another approach based on \textbf{parametric} model checking of Markov models.

\subsection{Parametric Model Checking}
%The projection of an FDTMC onto a product is a DTMC without features that represents the stochastic behaviour of that product. It means that only the features of the products have to be known to determine the actual probability values of the transitions. 
In this second method, we propose to convert an FDTMC into a parametric DTMC where the parameters are the features, and reuse existing methods for model checking parametric DTMCs. %In this subsection, we discuss how we can achieve this conversion. Since variability manifests itself in two forms, we show that each of these boils down to adding parameters to the Markov models.

%The first form of variability impacts on transitions availability. Let $\gamma(s,s')$ be the feature expression characterizing the availability of the transition from $s$ to $s'$. Then, for all products $p' \not\in \sem{\gamma(s,s')}$, $p'$ cannot execute the transition. In other words, the probability of the transition is 0 for these products. Remember that we define a product as a set of features. Since features can be regarded as Boolean variables, we can express the above property by transforming $\gamma(s,s')$ into an arithmetic parametric expression. A given assignation of features makes this expression equal to zero (resp. one) if and only if the corresponding product cannot (resp. can) execute the transition. A parametric expression  $\epsilon(e)$ is derived from a feature expression $e$ according to the following rules :
%$$\epsilon(e) = \left\lbrace 
%	\begin{array}{ll}
%		min(\epsilon(e_1), \epsilon(e_2)), & e = e_1 \land e_2,\\
%		max(\epsilon(e_1), \epsilon(e_2)), & e = e_1 \lor e_2,\\
%		1 - f, & e = \lnot f,\\
%		f, & e = f.
%	\end{array}
%\right .$$
%Once this parametric expression is derived, we replace $P(s,s')$ by $\epsilon(e) \times P(s,s')$. For a given product $p$ the execution probability of the transition is thus $0$ if $p \not\in \sem{\gamma(s,s')}$ and $P(s,s')$ otherwise. This corresponds exactly to Equation~\ref{eq:proj-p} in Definition~\ref{def:proj}. The transformation thus preserves the semantics of the original FDTMC.

In FDTMC, probability distributions are represented by the aforementioned probability profiles. We defined a probability profile as a function  that associates a product $p$ with a probability value. We can encode it as a parametric expression where the parameters are the features. For this purpose, we sum up the values that it can return, weighted by a parametric expression designating the corresponding products. We denote by $\epsilon(p)$ the parametric expression corresponding to the product $p$. Assume that the features of $d$ are boolean, i.e. can have a value either 0 (absent) or 1 (present), which is the most common case.
Let $\{f_1, \dots, f_k\} \in \sem{d}$ be the features of a product and $\{f'_1, \dots, f'_j\}$ the set of features that this product does not have. Then, $\epsilon(p)$ is given by
\begin{align}
	\epsilon(p) &= b_1 \times \dots \times b_k \times b_{k+1} \times \dots \times b_{k+j},\\
	b_i &= \left\lbrace 
	\begin{array}{ll}
		f_i, & i = 1 \dots k,\\
		1 - f'_{i-k}, & i = k+1 \dots k+j
	\end{array}
	\right .
\end{align}
This parametric expression is equal to 1 if we assign the value 1 to all the features of $p$ and the value 0 to all the others. For any other product, the expression is equal to 0. 

%The encoding of attributes is harder, as those may define a potentially infinite and uncountable number of different probability values. 

Using the above encoding, we represent the profile as the parametric expression
\begin{equation}
\label{eq:fparam}
	\sum_{p_i \in \sem{d}} \epsilon(p_i)\Pi(p_i).
\end{equation} 
When considering a particular product $p$, every term of this sum except one is equal to 0; the remaining term gives the probability. Note that if several products share the same probability value, we can drastically simplify the above sum. In particular, if the probability value depends only on a feature $f$, then we can rewrite Equation~\ref{eq:fparam} as $f \times \alpha_1 + (1 - f) \times \alpha_0$ where $\alpha_1$ (resp. $\alpha_0$) is the probability of $t$ when the feature $f$ is enabled (resp. disabled). %This way of reducing the complexity of parameters profile has already been illustrated in the previous subsection.

%The second form of variability makes the reward earned through the execution of a transition depend on features as well. To cope with it, we use a technique similar to the previous one, that is, we define the reward of a given transition as a parametric expression over the set of features. We thus represent $\Xi(s)$ by the expression
%\begin{equation}
%\label{eq:freward}
%	\sum_{p_i \in {\top, \bottom}^{|N|}} \epsilon(p_i) \times \Xi(s, p_i).
%\end{equation} 
%As for the previous case, we will often be able to write the above sum as a more concise expression.

By applying the above encoding, we reduce FDTMC verification to parametric DTMC verification, where the parameters model the presence or absence of boolean features. By doing so, we can benefit from efficient parametric model-checkers like PARAM~\cite{Hahn2010} and our tool~\cite{Filieri2011}. Given a parametric model, these tools return an expression containing parameters that encodes the probability we want to compute. To determine the actual probability that a given product satisfies the property, we replace, in the expression, each feature by 1 if it belongs to the product, or by 0 otherwise. The $\mathbb{P}$ operator  then yields a parametric inequality, that we need to transform in a boolean formula.
In the worst case, this can be done by  evaluating the expression once per valid product, at the cost of an exponential time.
% If we want to check that an expre can also solve the inequations

An advantage of this algorithm is that it performs only one exploration to compute the expression. However, %it either has to to determine the probability of satisfaction of the formula for all of these, or to solve the inequations expressing our requirements. Moreover, 
this expression becomes increasingly complex as the number of (feature-dependent) transitions to explore grows.

%\begin{figure}
%	\begin{center}
%		\includegraphics[scale=0.56]{figure/pMRM.pdf}
%	\end{center}
%	\caption{Parametric model corresponding to the wiper systems SPL.}
%	\label{fig:pDTMC}
%\end{figure} 
%
%\textbf{Example}. We applied this methodology to the aforementioned wiper systems SPL. We first transformed the FDTMC modelling this SPL into a parametric model. This operation yields the parametric DTMC shown in Figure~\ref{fig:pDTMC}. Then, we fed the model into the model checker of Filieri \textit{et al.}~\cite{Filieri2011}, which returned the rational expression
%$$\frac{\begin{split}
%				(-15*spd2*eco*very -15*spd2*eco + \\ 45*spd2*very +45*spd2 -40*eco +120)
%			\end{split}}
%		{8}.$$
%According to the above formula, the product with a very sensitive sensor and two speeds consumes 26.25 energy units on average. If the economic feature is enabled as well, this amount is reduced to 17.5 units.

\subsection{Feature-Aware Bounded Search}

Our third method relies on a novel algorithm that explores the FDTMC and keeps track of the variability and probability information obtained during the search. Its first step is to decompose the PCTL formula into its \emph{parse tree}, \textit{i.e.} a tree where each node is a state subformula of the original formula, such that the root is the formula itself, the leaves are atomic propositions, and child nodes form the direct subformulae of their parent. Similarly to CTL model checking algorithm for FTS~\cite{Classen2011}, we compute the satisfaction sets of all the subformulae %starting from the node and 
bottom-up. The satisfaction set of a state formula $\Phi$ is the set $Sat(\Phi) \subseteq S \times \sem{d}$ such that $(s,p) \in Sat(\Phi)$ iff $p$ satisfies $\Phi$ starting from $s_f$, noted $p \in \sem{(s \models \Phi)}$. Therefore, $s \models \Phi$ can be regarded as a Boolean formula that encodes the set of products for which $s$ satisfies $\Phi$. %The computation of PCTL satisfaction sets in FDTMCs is further complicated, as in addition to a set of products we also have to deal with the probability operator and the probability distribution of the transitions. 
The satisfaction rules of PCTL formulae are given as follows.

\begin{definition}
	Let $M$ be an FDTMC over an FD $d$, $s \in S$ one of its states. Then the satisfiability of a PCTL state formula by $s$ is calculated according to the following rules:
	\begin{equation*}
		\begin{array}{lcl}
		s,p \models \top &\Leftrightarrow & \top \\
		s,p \models a &\Leftrightarrow & a \in L(s) \\
		s,p \models \Phi_1 \land \Phi_2 &\Leftrightarrow & s,p\models \Phi_1 \land s,p \models \Phi_2\\
		s,p \models \lnot \Phi & \Leftrightarrow & \lnot (s,p \models \Phi) \\
		s,p \models \mathbb{P}_J (\Psi) & \Leftrightarrow & \Pi\big(s \models \Psi\big)(p) \in J
		\end{array}
	\end{equation*}
	where $\Pi(s \models \Psi)$ is a probability profile defined as $$\Pi(s \models \Psi)(p) = \Pi(\{ \rho \in Paths(s) \suchthat \rho, p \models \Psi\}).$$
	The satisfaction of path formulae is defined as follows:
	\begin{equation*}
		\begin{array}{lcl}
		\rho,p \models X\Phi &\Leftrightarrow & \rho[1],p \models \Phi\\
		\rho,p \models \Phi_1 U \Phi_2 &\Leftrightarrow & \exists j \suchthat (\rho[j],p \models \Phi_2 \land \\ & & \forall 0 \leq k < j \suchthat \rho[k],p \models \Phi_1)\\
		\rho,p \models \Phi_1 U^{\leq t} \Phi_2 &\Leftrightarrow & \exists j \leq t \suchthat (\rho[j],p \models \Phi_2) \land \\ & &\forall 0 \leq k < j \suchthat \rho[k],p \models \Phi_1)
		\end{array}
	\end{equation*}
	with $\rho = \rho[0], \rho[1], \dots$
\end{definition}
Since we have to use a probability profile for each state for computing the $\mathbb{P}_J$ operator, we also encode satisfaction sets this way,
using only 0 or 1 as results:
%as a set of probability profiles, \textit{i.e.} one for each state, such that the probability profile linked to state $s$ is
 $\textbf{1}_{s \models \Phi}$ with $$\textbf{1}_{s \models \Phi}(p) = \left \lbrace \begin{array}{ll} 1, & s,p \models \Phi \\ 0, & \text{otherwise.}\end{array} \right .$$ Apart from the probability operator, the computation of the satisfaction sets follows the same procedure as in~\cite{Classen2011}. A solution to determine for which products a state $s$ belongs to the satisfaction set $\mathbb{P}_J (\Psi)$ is to compute the probability profile $\Pi(s \models \Psi)$. If $\Psi = X \Phi$, then this profile is computed by:
$$\Pi(s \models X \Phi) = \sum_{s' \in S} \Pi(s,s') \otimes \textbf{1}_{s' \models \Phi}.$$
Indeed, for each product $p$, the probability that $s$ satisfies $X \Phi$ is equal to the probability that, in $p$, $s$ reaches a state satisfying $\Phi$ in one transition.

When $\Psi = \Phi_1 U^{\leq t} \Phi_2$, the probability profile is computed by solving the following recursive equations:
\begin{align}
\Pi(s \models \Phi_1 U^{\leq 0} \Phi_2) &= \textbf{1}_{s \models \Phi_2} \label{eq:puntil0}\\
\Pi(s \models \Phi_1 U^{\leq i} \Phi_2) &=\textbf{1}_{s \models \Phi_2} \oplus \\ &  (\textbf{1} - \textbf{1}_{s \models \Phi_2}) \otimes \textbf{1}_{s \models \Phi_1} \otimes \\ & \sum_{s' \in S} \big( \Pi(s,s') \otimes \Pi(s' \models \Phi_1 U^{\leq i-1} \Phi_2)\big) \label{eq:puntil}
\end{align}
where $i > 0$. Indeed, according to the until operator, the probability that in a product $p$, $s$ satisfies $\Phi_1 U \Phi_2$ in zero step is 1 if $s$ satisfies $\Phi_2$, and 0 otherwise. The probability that $s$ satisfies $\Phi_1 U \Phi_2$ in $i > 0$ steps in $p$ is the probability that it satisfies the formula  $0$ steps or in $j$ steps, with $0 < j \leq i$. To satisfy the formula in $j$ steps in $p$, $s$ must (1) not satisfy $\Phi_2$ in $p$ (otherwise it satisfies the formula in zero step), (2) satisfy $\Phi_1$ in $p$ (otherwise it cannot satisfy the formula at all), and (3) reach a direct successor $s'$ that satisfies the formula in at most $i-1$ steps.

The correctness of the above equations directly follow from the semantics of the bounded until, the definition of projection, and the expansion laws of the bounded until operator~\cite{Baier2007}. In the case of unbounded until, the probability value is 
classically obtained by removing the superscripts in the equations and solving the resulting system of linear equations \cite{Baier2007}.
But we can also obtain the solution by iterating these equations for an increasingly high bound $k$. %as the limit. 
We obtain lower approximations of the desired probability values, that increase with $k$ and tend to the exact value. % with increasing $k$. %The accuracy of this approximation increases with the bound.
%If we only want to know whether the probability is above a given threshold $\mathbb{P}_{> p}$, the computation can be abandoned as soon as the threshold is reached. % for all products.
%only if this is true for all states it can influence other states

We iterate these equations by performing a bounded exploration on the FDTMC. Algorithm~\ref{alg:fbe} presents this exploration procedure. Following the principles of FTS model checking~\cite{Classen2013}, the algorithm ensures that a given path is visited only once. The idea of this algorithm  is to start from the set of states that satisfy $\Phi_2$ for at least one product. Then we perform a backward exploration to discover new paths that satisfy $\Phi_1 U^{\leq k} \Phi_2$. % as well as the probability that each product executes these paths.
For each state found along these paths, we record 
%the products and the probability with which the state satisfies the until formula. 
%At the end, we obtain for each state $s$ a probability profile that is an upper approximation of $\Phi_1 U^{\leq k} \Phi_2$ and a lower approximation of $\Pi(s \models \Phi_1 U \Phi_2)$.
%
%The algorithm behaves as follows. For each state, we maintain
%\begin{itemize}
%\item  
a variable $x_s$, a probability profile
that is a lower approximation of $\Pi(s \models \Phi_1 U \Phi_2) $ and will eventually reach a value above $\Pi(s \models \Phi_1 U^{\leq k} \Phi_2) $; therefore it tends to $\Pi(s \models \Phi_1 U \Phi_2) $ with higher $k$.
%\item a set of states $N_s$, the already discovered successors of $s$ along a path satisfying $\Pi(s \models \Phi_1 U \Phi_2) > \bf(0)$.
%\end{itemize}

First, for each state $s$, we record the set of products for which $s$ satisfies $\Phi_2$ (Line 3). As explained above, this can be encoded as the probability profile $\textbf{1}_{s \models \Phi_2}$. If there is at least one such product, for every possible predecessor $s'$ of $s$ that can reach one of those products, we push the transition from from $s'$ to $s$ together with a number $1$ that indicates that the algorithm analyses this transition as part of a path of length 1 (Lines 4--6). Next, we iteratively analyse the transitions on the stack in order to explore paths of greater lengths. Let $(s,i,s')$ be the top element of the stack (Line 9). If $i$ exceeds $k$, we skip the element since our approximation bound is reached.
%; otherwise we would analyse 
%it corresponds to a path of length greater than the specified bound $k$. 
Then we apply the equations %of the bounded until 
to recalculate $x_s$ using the new probability values found for $s$ %(and the old values for the other successors)
 (Line 11). Note that by doing so, we might go above $\Pi(s \models \Phi_1 U^{\leq k} \Phi_2) $ since $x_{s_u}$ might already include paths longer than the current path, giving thus an even better approximation of $\Pi(s \models \Phi_1 U \Phi_2) $. 
 If one of the values changed, we may have to update the value of every predecessor of $s$. To that aim, we add a new triplet $(s'',i+1,s')$ on stack where $s''$ is a predecessor of $s'$. 
\begin{algorithm}
	\caption{Feature-aware bounded search}
	\label{alg:fbe}
	\begin{algorithmic}[1]
		\REQUIRE An FDTMC, two PCTL state formulae $\Phi_1$ and $\Phi_2$, an integer bound $k \geq 0$.
		\ENSURE For each $s \in S$, $\Pi(s \models \Phi_1 U^{\leq k} \Phi_2)$. 
		%\STATE $\Pi(\Phi_2) : S \rightarrow (\sem{d} \rightarrow [0,1]) : \Pi(\Phi_2)(s) = \textbf{1}_{s \models \Phi_2}$;
		%\STATE $Sat(\Phi_1 U \Phi_2) : S \rightarrow (\sem{d} \rightarrow [0,1]) : Sat(\Phi_1 U \Phi_2)(s,p) = \left \lbrace \begin{array}{ll} 1, & Sat(\Phi_2, s, p)  \\ 0, & \text{otherwise}\end{array} \right .$;
		\STATE $Stack \leftarrow []$;
		\FOR{$s' \suchthat \textbf{1}_{s \models \Phi_2} \neq \textbf{0}$}
			\STATE %$\Pi(s \models \Phi_1 U^{\leq k} \Phi_2)
                        $x_{s'} \leftarrow \textbf{1}_{s' \models \Phi_2}$;
			\FOR{$s \suchthat \Pi(s,s') \otimes \textbf{1}_{s' \models \Phi_2} \neq \textbf{0}$}
				\STATE $Stack \leftarrow push(Stack, (s, 1, s'))$;
			\ENDFOR
		\ENDFOR
		\WHILE{$Stack \neq []$}
			\STATE $(s, i, s') \leftarrow pop(Stack)$;
			\IF{$i \leq k$}
                                %\STATE $N_{s}    \leftarrow  N_{s}  \union \{s'\}$
				\STATE $new \leftarrow \textbf{1}_{s \models \Phi_2} \oplus (\textbf{1} - \textbf{1}_{s \models \Phi_2}) \otimes \textbf{1}_{s \models \Phi_1} \otimes \sum_{s_u \in S} \Pi(s,s_u) \otimes x_{s_u}$;
				\IF{$new > x_s$}
					\STATE $x_s \leftarrow new$;
					\FOR{$s'' \suchthat \Pi(s'',s) \neq \textbf{0}$}
						\STATE $Stack \leftarrow push(Stack, (s'', i+1, s'))$;
					\ENDFOR				 
				\ENDIF
			\ENDIF
		\ENDWHILE
		\RETURN $x$
	\end{algorithmic}
\end{algorithm}
The algorithm always terminates since there is a finite of paths bounded by $k$. The correctness is ensured by Equations~(\ref{eq:puntil0}) and~(\ref{eq:puntil}). 
%It is also possible to design a variant where a path is abandoned when the change of value is small enough.

For the bounded until, the algorithm is similar, but it works in $k$ phases. %, with a phase for each $i \leq k$. 
It uses two profiles: $x_s$  contains the previous iteration: $x_s = \Pi(s \models \Phi_1 U^{\leq i} \Phi_2) $, and computes a new profile  $x'_s = \Pi(s \models \Phi_1 U^{\leq i+1} \Phi_2) $. The ``stack'' is first emptied of $i$-edges before dealing with $i+1$ edges, which amounts to a breadth-first search.

The advantage of this bounded method is that it checks all the products in one exploration.
% and avoids redundant verification of the same execution path. 
Unlike the parametric method, our feature-aware search does not require to evaluate a rational  expression for each of the products. Instead, for each PCTL state formula, it returns a Boolean formula encoding which products satisfy it. % in the initial state $s_0$.

\section{Experiments}
\label{sec:validation}

In this section, we report the results obtained by evaluating the performance of the three FDTMC verification techniques in terms of verification time. We consider two technical case studies as our benchmarks, which we systematically extend to obtain larger models. All the models are available on \url{http://info.fundp.ac.be/~pys/fdtmc/}.

 %The first verification approach consists in deriving the MRM of each individual product and then performing the conventional probabilistic model checking. The second approach applies parametric model checking techniques to verify a stochastic SPL specified by FMRM. The result of parametric verification is a rational function where variables are the features of the FMRM. Given an individual product, the rational function is evaluated by replacing each variable by 1 if the corresponding feature is part of the product, and 0 otherwise. Regarding the properties, we focus on PCTL.%reliability and energy consumption as two important non-functional requirements. 

%\subsection{Empirical Results}
%\label{sec:emp}

%To the best of our knowledge there is no repository of SPL specifications that can be used as the benchmark for (stochastic) model checking and verification purposes. Thus, 

The first case study is an abstract model of failure-prone systems. In this model, the system has to go through successive degradation states to eventually reach an absorbing failure state. In every degradation state, however, instead of going to the next degradation state, the system may completely break and reach a second absorbing failure state. In every degradation state, the system may also partially recover and reach the previous degradation state. Apart from the initial state and the two absorbing states, the probability of the transitions leaving each state depends on the presence of absence of specific features. The model is extended by adding new degradation states and features. 

The second case study is an abstract model of a service provider system that gives the opportunity to its users to invoke different services. The execution of a service is modelled by a sequence of states. During such executions, the system may fail and suddenly reach a failure (absorbing) state. After any service execution, the system may keep executing more services or may go to an absorbing successful-termination node. Each service requires a specific feature to be started, hence the variability within such a system. Unlike the first model, the behaviour of the features are completely independent in this model. We enlarge the model by gradually adding new services, which also increases the number of features. 

For both examples, we checked that the probability that system reaches a failure state is below 0.1. This reachability property can be expressed in PCTL as $\mathbb{P}_{< 0.1} \big(\diamond failure)\big)$. All benchmarks were run on a Dual Intel(R) Xeon(R) CPU E5530 @2.40GHz with 8Gb of RAM, equipped with GNU Linux Ubuntu server 11.04 64bit. To perform the enumerative verification, all the DTMCs modelling a specific product are first derived from the FDTMC and then verified one-by-one by using the PRISM model checker\footnote{http://www.prismmodelchecker.org/}~\cite{Kwiatkowska2009}. For the parametric approach, we use the parametric model checker developed by Filieri \textit{et al.}~\cite{Filieri2011}, and then evaluate the resulting expression by using JEP Java library\footnote{\url{http://www.singularsys.com/jep/}}. For the bounded approach, we use a prototype we developed from scratch. The latter is available on \url{http://info.fundp.ac.be/~pys/fdtmc/} as well.

The total verification time of the enumerative approach is the sum of the model-checking times to verify each single product by PRISM. We excluded the time to produce individual DTMCs as well as the time taken for creating PRISM input files. As for the parametric approach, the verification time is obtained by summing the parametric verification time and the time spent to evaluate the expression for each single product. Since the bounded approach verifies all products together, its verification time equals to the total time taken by the prototype tool. In all the experiments, we set the bound of the algorithm such that the maximum precision error is always less than $10^{-3}$.

\begin{figure}
	\begin{center}
		\includegraphics[scale=0.46]{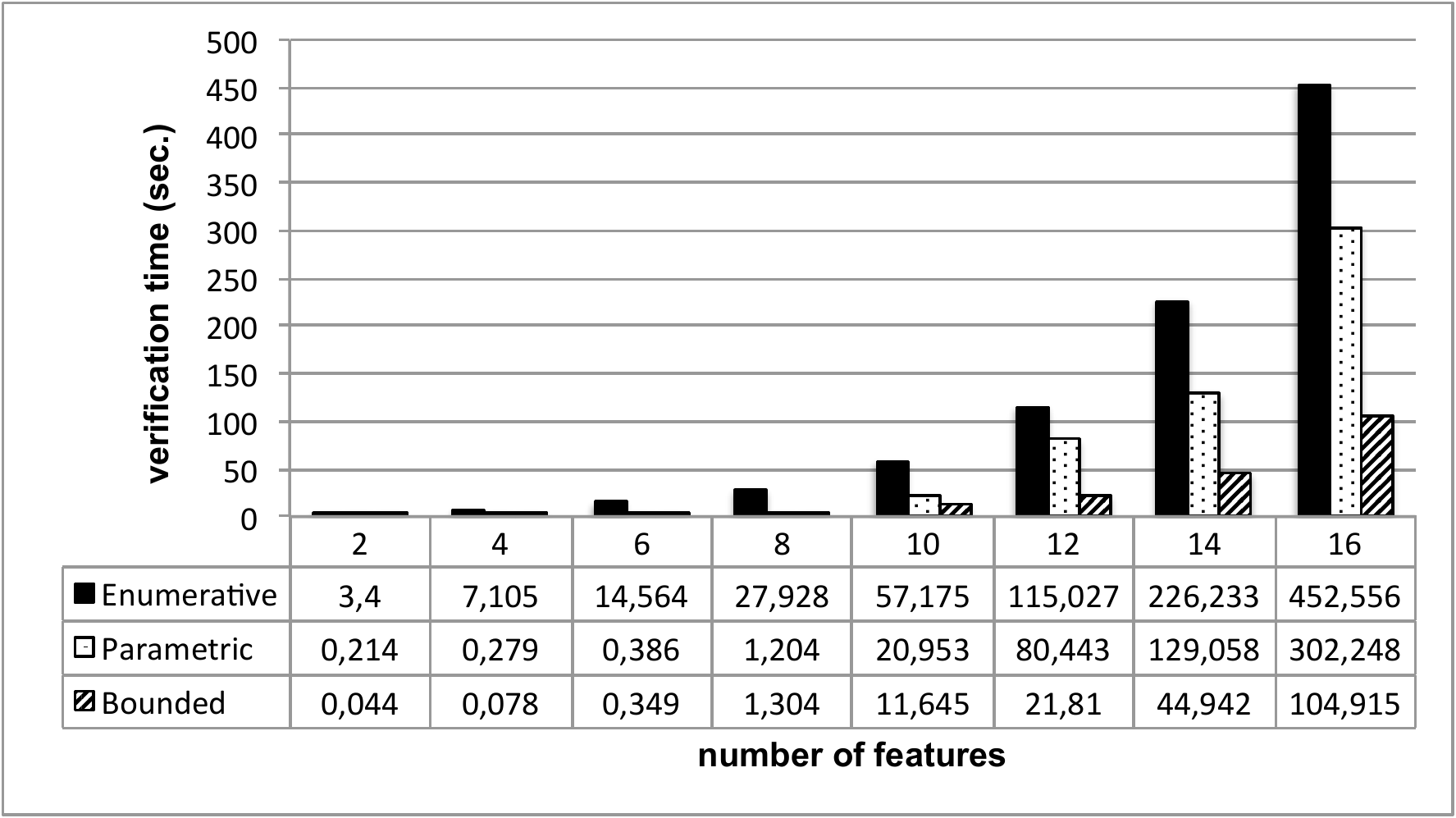}
	\end{center}
	\caption{The verification time of failure-recovery case study}
	\label{fig:case1}
\end{figure}

Figure \ref{fig:case1} shows the verification times for the failure-recovery case study. In this case, the number of features $f$ grows from 2 to 16. It turns out that the bounded approach outperforms the others in almost every case. The verification time of the enumerative approach grows exponentially with the number of features, as expected. We also observed that the parametric approach suffers from the growing complexity of the rational function.

Table \ref{tb:case2} reports the time each of the verification techniques takes to verify the models of the service provider case study. The results show that the enumerative approach takes a longer time, while the other two techniques exhibit a similar performance. On contrary to the first case, the parametric approach outperforms the bounded technique.

\begin{table}[h]
\caption{The verification time of service provider case study (in seconds)}
\begin{center}
\begin{tabular}{| c | c | c | c |}
\hline
 Features & Enumerative & Parametric & Bounded \\ \hline
 2 & 4,207 & 0,237 & 0,111\\ \hline
 4 & 6,932 & 0,267 & 0,306\\ \hline
 6 & 14,57 & 0,336 & 0,314\\ \hline
 8 & 29,224 & 0,408 & 0,519 \\ \hline
 10 & 57,215 & 0,53 & 0,676\\ \hline
 12 & 119,544 & 0,572 & 1,636\\ \hline
 14 & 227,91 & 0,99 & 1,931\\ \hline
 16 & 466,185 & 1,126 & 2,966\\ \hline 
\end{tabular}
\end{center}
\label{tb:case2}
\end{table}

%\subsection{Discussion}
%\label{sec:discussion}

The results of our experiments suggest that the enumerative approach is increasingly inefficient as the number of features (and hence of products) increases. Still, if a verification task only deals with a small number of products, the enumerative approach is a reasonable choice. The cost of the parametric approach is highly dependent on the complexity of the rational function that is produced by the parametric model checker. This complexity varies depending on the topology of the verified model. In the first case, where feature-dependent transitions occur sequentially, the verification time grows faster than in the second case, where the features are scattered around the model. Our third algorithm exhibits a good performance in both experiments. In the second case, it remains competitive in spite of the high efficiency of the parametric approach.

Our theory is that the parametric approach performs better in models where feature-dependent transitions do often not occur in sequence, that is, when there is a limited number of \emph{feature interactions}. On the contrary, if many of these sequences occur in the model then the size of the  function returned by the parametric algorithm will sharply grow. In such cases, our feature-aware bounded search should instead be used.

\section{Related Work}
\label{sec:related-work}

Analyzing non-functional properties of software systems has received an increasing interest during the last years. However, there are only a few work discussing this issue for SPLs ~\cite{Bartholdt2009, Leire2008}. The recent research carried out addressed the problem at feature models or available source codes. Our approach instead focuses on the use of behavioural models. Feature models are suitable to specify and organize variability of SPLs, but they are not enough expressive and precise for quality analysis. On the other hand, analyzing the source code of an SPL is only possible after the implementation, and is not applicable at the early stage of development ~\cite{Leire2008}. Ghezzi and Molzam Sharifloo~\cite{Ghezzi2011}, as well as Nunes~\cite{Nunes2012} propose to use parametric model checking to check PCTL formulae on all the variants of an SPL. Yet, their modelling formalisms do not include an explicit notion of features and they do not propose alternative verification algorithms.

Given the increasing popularity of product lines in various areas including critical systems, SPL verification methods are actively studied, although most of them do not consider non-functional properties. The work surrounding FTS is the most related to ours~\cite{Classen2010,Classen2011,Classen2013,Cordy2013}. % Classen \textit{et al.}~\cite{Classen2010} introduce a first definition of this formalism and provide an efficient LTL model-checking algorithm. In~\cite{Classen2011,Classen2013}, they provide a new definition which uses feature expressions to model variability, and study CTL model checking. 
Several alternative to FTS exist. Larsen \textit{et al.}~\cite{Larsen2007} show that I/O automata are convenient for modelling product lines as open systems. Asirelli \textit{et al.}~\cite{Asirelli2011} equip modal transition systems with a logic able to
express constraints on variable behaviour. Gruler \textit{et al.}~\cite{Gruler2008} extend the process algebra CCS with variability operators, which allow to model alternative choices between two processes. Li \textit{et al.}~\cite{Li2002b} model both the base systems and optional features as finite state machines that connect to each other. Apel \textit{et al.}~\cite{Apel2011} specify features as separate units that can be composed; each feature defines safety properties that are subsequently verified using single-system model-checkers.

Outside the context of software product lines, we find the notions of Interval Markov chain~\cite{Jonsson1991} and Constraint Markov chain~\cite{Caillaud2010}. The former is a generalization of Markov chains where execution probability of transitions are given by probability intervals rather than constant values. Interval Markov chains thus concisely models a potentially infinite set of Markov chains. Then, it is possible to determine whether or not all those Markov chains satisfy a given PCTL property. Constraint Markov chains are a generalization of Interval Markov chains, where probability distributions are defined by parameters. The values that a parameter can take are determined by linear constraints between the parameters (e.g. $\alpha + \beta = 0.7$). This kind of parameterization is more general than that of FDTMC with Boolean features as presented in this paper. By extending FDTMC with numeric features~\cite{Cordy2013}, we obtain a formalism as expressive as Constraint Markov chains. Moreover, to the best of our knowledge there exists no algorithm for checking Constraint Markov chains against properties expressed in PCTL.

\section{Conclusion}
\label{Conclusion}

In this paper, we tackled the verification of non-functional requirements in software product lines. We extended the probability theory with notions of variability, and applied this extension to define DTMCs enriched with variability operators, aka FDTMCs. We showed that when the inner behavior of a software is not stochastic, we can model a stochastic SPL as the composition of an FTS modelling the system and a set of DTMCs modelling the stochasticity of the environment. We discussed and experimented three methods to model check FDTMCs, including a novel algorithm that directly exploits variability information contained in the model. Benchmarks suggest that enumerative verifications of the whole product line take a long time w.r.t. the other two methods. 

In the future, we plan to develop a complete tool chain that will provide assistance in both the specification and verification of stochastic product lines. We will support specification of FDTMCs either via a graphical user interface or through the use of high-level textual languages. Moreover, our tool will make use of the compositional modeling approach presented in this paper. To achieve the verification, we plan to implement and optimize our feature-aware bounded algorithm within the tool and will link parametric model checkers to the tool.

A natural direction for our future work is the extension of the proposed approach to other kinds of Markov models, for example continuous time Markov chains, and Markov reward models, which are widely used to reason about other kinds of non-functional requirements such as performance and energy-consumption. The enrichment of Markov processes as presented in Section~\ref{sec:probability} provides a formal framework from which these new models can be derived. Moreover, our compositional modelling method can be applied to these models as well. Still, the major challenge remains the design of efficient verification algorithms.

Also, we want to deepen the current approach to answer optimizations problems. Nowadays, optimizing non-functional requirements is an important challenge in a wide range of areas. For instance, we could consider the problem of finding the most reliable and economic products among a product line. Since an FDTMC can be regarded as a parametric model with Boolean variables (namely the features) only, addressing this problem leads us in the particular domain of Boolean linear optimization. As an alternative to model checking and algebraic methods, we will also investigate simulation-based methods for computing probabilities and rewards~\cite{Younes2002}. %Applying these methods in the context of product lines is a challenge because in order to be efficient, a simulation must benefit from the commonalities between the products. We thus need a method for splitting a simulation between several parts of a given model. %Additionally, simulations can be appropriate for figuring out the optimal products. 

%------------------------------------------------------------------------- 
%\section*{Acknowledgments}
%This work was partially funded by the Belgian Fund for Scientific Research - FNRS; and European Commission, Programme IDEAS-ERC, Project 227977-SMScom.
\bibliographystyle{plain}
\bibliography{mcr}

\end{document}